\documentclass[epj]{svjour}
\usepackage{graphics}
\begin{document}
\title{Pygmy resonances and symmetry energy}
\author{C. A. Bertulani
}                     
\offprints{}          
\institute{Department of Physics and Astronomy, Texas A\&M University-Commerce, Commerce, TX 75429, USA}
\date{Received: date / Revised version: date}
%
\abstract{
I present a brief summary of the first three decades of studies of pygmy resonances  in nuclei and their relation to the symmetry energy of nuclear matter. I discuss the first experiments and theories dedicated to study the electromagnetic response in halo nuclei and how a low energy peak was initially identified as a candidate for the pygmy resonance. This is followed by the description of a collective state in medium heavy and heavy nuclei which was definitely identified as a pygmy resonance. The role of the slope parameter of the symmetry energy in determining the properties of neutron stars is stressed. The  theoretical and experimental  information collected on pygmy resonances, neutron skins, and the numerous correlations found with the slope parameter is  briefly reviewed.
} 
\maketitle

{\it To the memory of Pier Francesco Bortignon}

\bigskip
\section{Introduction}
\subsection{Giant Resonances}
\label{Giants}
Collective excitation modes in nuclei are well known and have been studied for the last 70 years. Their first unambiguous observation was reported by Baldwin and Klaiber in  \cite{BK47}  in photo-absorption experiments. But Bothe and Gentner \cite{BG37} already had a first indication that they existed when they obtained very large cross sections, two orders of magnitude larger than predicted theoretically, for the photo-production of radioactivity in several targets. The large cross sections were obtained with high energy photons, of about 15 MeV, and now are understood as due to a collective nuclear response to the  electric dipole (E1) field of the photon. Such E1 collective states had been predicted theoretically by Migdal \cite{Mig44}. For a detailed description of giant resonances, see Ref. \cite{BBB98}

Much of what we know about the giant resonances today was gathered in photo-absorption processes using mono-energetic photons, as reported, e.g., in Refs.  \cite{BF75,Ber77}. The giant resonances  in all nuclei  are located above the particle emission threshold.  In medium and heavy nuclei the large Coulomb barrier prevents charged particle decay, and the photo-absorption cross section is usually obtained from a measurement of neutron yields for a given $\gamma$-ray energy \cite{Aum05,ABS95,BV99}. 

\subsection{Origins of the Pygmy Resonance}
\label{Origins}

Giant resonances are observed in basically all nuclei. In contrast, pygmy resonances have been mainly observed in neutron-rich nuclei. Historically, the first observation of a pygmy resonance occurred in 1961 with the discovery of  a significant number of unbound states identified as a bump in $\gamma$-rays emitted following neutron capture \cite{Bar61}. But the first use of of the name {\it pygmy resonance}, or pygmy dipole resonance (PDR), was in 1969 when its effect on the calculations of  neutron capture cross sections was reported \cite{Brz69}. \footnote{I thank Riccardo Raabe (KU Leuven) for this information.}

The description of the PDR as a collective excitation was first introduced in Ref. \cite{MDB71} based on a  three fluid model consisting of protons and neutrons fluids in the same orbitals, and neutrons accounting for excess neutrons interacting less strongly with the other nucleons. The model assumed that the neutron excess would oscillate against the $N = Z$ core. It was much later when the first experimental proposal to study pygmy resonances in Coulomb excitation experiments emerged in 1987 at the JPARC facility in Japan \cite{NK87}.

\section{Low Energy Response in Neutron-Rich Nuclei}
\subsection{Nucleon Knockout Reactions}
\label{Binding}
The abnormal size of the $^{11}$Li nucleus was first reported in Ref. \cite{Tan85} by measuring the momentum distributions of $^9$Li fragments in nucleon removal reactions. A superposition of two nearly gaussian shaped distributions was necessary to reproduce the experimental data. The wider peak was linked to the removal of neutrons from a tightly bound $^{9}$Li core, whereas the narrower peak was thought to arise from the removal of the loosely bound valence neutrons in $^{11}$Li. In fact, the two-neutron separation energy in $^{11}$Li is only about  300 keV explaining why their removal only slightly ``shakes'' the $^{9}$Li core, thus explaining the narrow component of the momentum distribution. A narrow momentum distribution implies a large spatial extent of the neutrons, which in turn is a consequence of their small separation energy. These conclusions were later found to be in agreement with Coulomb breakup experiments \cite{Kob89} and with theory \cite{HJ87,BB86}. 

The momentum distributions of core fragments of halo projectiles were initially analyzed experimentally using the simple Serber formula \cite{Ser47}
\begin{equation}
{d \sigma_{c}\over d^{3} q} ={\cal C} \left| \psi({\bf q})\right|^{2},
\end{equation}
where $\cal C$ is a kinematical constant and $\psi({\bf q})$ is the Fourier transform of the ground state wave function of the nucleus. In fact, this formalism works rather well for loosely-bound nuclei such as $^{11}$Li and $^{11}$Be \cite{Orr92,Aum00,BH04}. For nucleon removal with larger separation energies, the Serber formalism is not appropriate and yields inaccurate results \cite{BH04,BM93,HBE96,BH00,Mar02,BCH93}.

\subsection{Electromagnetic Response}

A narrow peak associated with the small separation energies was also observed in Coulomb dissociation experiments \cite{Iek13,Sac13}, becoming the seed for subsequent intensive investigations. 

The Coulomb excitation cross section for a given multipolarity $\pi L$ ($\pi = E$ or $M$, and $L=1,2,\cdots$) and excitation energy $E$ is given by 
\begin{equation}
{d\sigma \over dE} = {n_{\pi L}(E)\over E} \sigma^{\pi L}_{\gamma}(E), \label{dsde}
\end{equation}
where $n_{\pi L}$ is the virtual photon number \cite{BB88} and  $\sigma^{\pi L}_{\gamma}$ is the photo-nuclear cross section, which can be written in terms of the electromagnetic response function $dB_{\pi L}/dE$ as \cite{BB88}
\begin{equation}
\sigma^{\pi L}_{\gamma}(E)= {(2\pi)^{3}(L+1) \over L [(2L+1)!!]^{2}} \left( {E\over \hbar c} \right)^{2L-1}{dB_{\pi L}(E) \over dE} \label{photo}.
\end{equation}

\subsection{Low Binding Energy Effect}
\subsubsection{Two Body Models}
Using simple Yukawa or Hulthen functions for bound states and plane waves for the continuum  it was shown in Refs. \cite{BB86,BBH91,BS92} that  a simple expression for the response function emerges for electric multipoles, namely,  
\begin{eqnarray}
{dB_{E L}(E) \over dE}&=& {2^{L-1} \over \pi^{2}} (2L+1)(L!)^{2}\left( { \hbar^{2} \over \mu} \right)^{L} \nonumber \\
&\times& Z^{2}e_{L}^{2}{\sqrt{S} (E-S)^{L+1/2}\over E^{2L+2}},\label{dbde1}
\end{eqnarray}
where $e_{L}$ is the effective charge, $S$ is the separation energy and $\mu$ the reduced mass \cite{BB86,BBH91,BS92}. 
For the ubiquitous electric dipole (E1) excitation one has
\begin{eqnarray}
{dB_{E 1}(E) \over dE}&=& {3 Z^{2}e_{L}^{2}\hbar^{2}\over \mu \pi^{2}}  {\sqrt{S} (E-S)^{3/2}\over E^4}.\label{dbde2}
\end{eqnarray}
These equations are very useful as they allow simple predictions of the Coulomb response in halo nuclei in terms of the separation energy $S$. They have been widely used in experimental analyses \cite{Nak94,Shi95} and for comparison with more complex theoretical models \cite{BS92,Ter91,Ots94,Barr01,Pot10,Barr17,Man19,Brog19,Barr19}. The above equations predict a peak in the response function at an excitation energy of $E=8S/5$ and a width approximately equal to $E$ \cite{BS92}. 

A certain confusion reigned in the literature during the 1990s as to whether the peak observed in the Coulomb breakup experiments was due to a low energy resonance or just a direct transition to the continuum, as is the case behind the derivation of Eqs. \ref{dbde1} and \ref{dbde2} \cite{BS92}.  

Despite the appeal and usefulness of the expressions above, it was later on realized that in order to reproduce many of the experimental data gathered on Coulomb breakup of halo nuclei during the 1990s and 2000s it was necessary to incorporate higher-order interactions, or final state interactions (FSI), in the theoretical calculations. The discretization of the continuum was a crucial improvement of the theories and the so-called continuum-discretized-coupled-channels (CDCC) calculations became a necessary theory method to reproduce experimental data and to probe the structure of loosely bound states and resonances appearing in halo systems. The role of higher-order couplings and the nuclear contribution to the breakup and comparison to experimental data has been studied by numerous authors \cite{BC92,BB93,BS93,Kid94,BS96,Kid96,Ban02,Mar03,IB05,Bert05,TB05,Mor05,OB09,Oga09,OB10,KB12,HTS13,Shub15,Ala17,BBFV16}. Calculations based on effective field theories started to emerge only recently \cite{Cap18}.

\subsubsection{Three Body Models}
Three body models for halo nuclei such as $^{11}$Li and $^{6}$He have obtained a similar response function as described in the previous section \cite{PJZ96,Zhu93,Dan98,Ber07}, yielding  ${dB_{E 1}(E) / dE} \propto (E-S)^{3}/E^{11/2}$. This model shifts the three-body response peak to  larger excitation energy $E$ than the two body model. Therefore, the separation energy still roughly determines the peak location of the E1 response but at a higher energy and with a larger width. It was also shown that final state interactions can substantially change the location of the low energy peak \cite{Ber07,TB05}.

Three-body models did not deliver much more information on the origins of the low-peak response and its identification as a pygmy resonance. The conclusion reached by using these models was that the observed peak was again due to a direct transition from the ground state of the nucleus, e.g., $^{11}$Li or $^{6}$He, to the continuum \cite{PJZ96,Zhu93,Dan98,Ber07}. This does no mean however that the three-body models were not rich in physics details and predictions, much on the contrary. They have helped us to understand new phenomena in ``Borromean'' nuclei and the unravelling of Efimov states in nuclei \cite{Zhu93}.   

\subsubsection{Hydrodynamical Models}
As the evidence accumulated of a non-negligible strength in the low energy spectrum of dipole excitations, also for heavier nuclei, other models for the pygmy resonances emerged \cite{MDB71,SIS90,VNW92}. It was natural to extend  the Goldhaber-Teller (GT) \cite{GT48} and  Steinwedel-Jensen (SJ) \cite{SJ50} hydrodynamical models to explain pygmy resonances as a collective response to external fields in ``soft'' neutron-rich nuclei. 

Using the concepts described in Ref. \cite{Mye77} one can show that the radial transition density of pygmy resonances can be described in the hydrodynamical model by the equation \cite{Ber07}
\begin{eqnarray}
\delta \rho(r) = \sqrt{4\pi \over 3} R \left[Z_{\scriptscriptstyle GT}\alpha_{\scriptscriptstyle GT} {d\over dr} + Z_{\scriptscriptstyle SJ}\alpha_{\scriptscriptstyle SJ}{K\over R} j_{1}(kr)\right]\rho_{0}(r) , \label{GTSJ1}
\end{eqnarray}
where $Z_{i}$ are the effective charges  in the GT and SJ models \cite{Ber07}, $\alpha_{i}$ are admixture coefficients of the GT and SJ collective vibration modes, such that $\alpha_{\scriptscriptstyle GT}+\alpha_{\scriptscriptstyle SJ}=1$, $K=9.93$ and $R$ is the mean nuclear radius. Here, $j_{1}(kr)$ is the spherical Bessel function of order 1, and $k=2.081/R$. The transition density in Eq. \ref{GTSJ1} emulates a pygmy resonance based on a collective dipole vibration of protons and neutrons and is known as {\it soft dipole mode}. 

If one choses $\alpha_{SJ}=0$, Goldhaber and Teller \cite{GT48} gave a simple prescription for the resonance energy of collective vibrations in nuclei,
 \begin{eqnarray}
E_{R}= \left({3 {\cal S} \hbar^{2}\over 2 aRm_{N}}\right)^{{1/2}}, \label{GTSJ}
\end{eqnarray}
where $a$ is the approximate size of the nuclear distribution thickness and $m_{N}$  is the nucleon mass. According to  Goldhaber and Teller \cite{GT48}, ${\cal S}$ in the equation above is not the separation energy of a nucleon but the energy needed to extract one proton from the neutron environment (or one neutron from the proton environment). It is the part of the potential energy due to the neutron-proton interaction in the nuclear environment, assumed to be proportional to the {\it symmetry energy} $\propto (N-Z)/A$.  

Golhaber and Teller \cite{GT48}  used ${\cal S} = 40$ MeV, $a=1-2$ fm and got $E\simeq 10-20$ MeV for a medium heavy nucleus, consistent with the experimentally found centroids of giant dipole resonances.  For neutron rich nuclei, the extension of this theory needs to include a relation of $\cal S$ to the symmetry energy. Such a relationship is only possible with a microscopic model due to the nature of the fine-structure of the pygmy resonance closely related to the coupling of phonon states with complex configurations in the nucleus. In the case of halo nuclei a hydrodynamical model is likely unfit, but it works if one assumes  that ${\cal S} = S$ (here the separation energy). For halo nuclei the product $aR$ is also expected to be proportional to $S^{-1}$, and we obtain the proportionality $E_{PDR}\sim \beta S$, with $\beta$ of the order of one . Using for example $^{11}$Li, with $a=1-2$ fm, $R=3$ fm, and $S = 0.3$ MeV, one gets $E_{PDR} = 1-2$ MeV, which is also compatible with experimental results.  

Evidently, the hydrodynamical models are useful to understand the physics of collective vibrations, but they lack accuracy. Microscopic models starting from a nucleon-nucleon interaction often relying on the linear response theory, e.g., the random phase approximation, is a better approach to describe giant resonances. Such models have also been used to study the fine-structure of the low-energy response in nuclei \cite{Barr01,Pot10,Barr17,Brog19,Barr19,Pon19}.

\subsubsection{Microscopic Models}

The random phase approximation (RPA) is a useful tool to describe the nuclear response function in terms of microscopic degrees of freedom. In its simplest form, one can calculate the response to a weak time-dependent field of the form $V_{x}({\bf r}) \cos(\omega t)$ by solving the RPA equations in the self-consistent method \cite{SG75} 
 \begin{eqnarray}
\delta \rho_{{RPA}} ({\bf r}) = \int \Pi^{{RPA}} ({\bf r, r'}) V_{x}({\bf r'})d^{3}r',
\end{eqnarray}
where $\delta \rho_{{RPA}}({\bf r})$ is the self-consistent transition density and $\Pi^{{RPA}} ({\bf r, r'}) $ satisfies the implicit equation
 \begin{eqnarray}
\Pi^{{RPA}} ({\bf r, r'}) &=&\Pi^{{0}} ({\bf r, r'}) + \int d^{3}r_{2}d^{3}r_{3 } \nonumber \\
&\times& \Pi^{{0}} ({\bf r, r}_{2}){\delta V_{x}({\bf r}_{2})\over \delta \rho({\bf r}_{3})}\Pi^{{RPA}} ({\bf r}_{3}, {\bf r'}) .
\end{eqnarray}
Here,  $\rho({\bf r})$ and $\Pi^{{0}} ({\bf r, r'})$ are the corresponding densities and response function defined in terms of occupied and unoccupied orbitals in the nucleus. Other variations of the RPA exist and are explained in details elsewhere, for example using the XY-formalism \cite{RS04,Col13} (for a recent review, see Ref. \cite{Brog19}).

The first application of the RPA formalism to obtain the electromagnetic response in weakly-bound nuclei was done in Ref. \cite{FB90}. This was followed up in Ref. \cite{BS92,Ter91} where a comparison with two-body models and nucleon clustering was done. In the RPA calculations a peak at small energies appears around a few MeV which was interpreted as a pygmy resonance, or alternatively just the effect of a small separation energy in the nuclei and of a direct transition to the continuum. At least for light nuclei. But for medium heavy and heavy nuclei, a fine structure of the resonance was revealed leading to the belief that many nucleons and many states are involved in the transition. 

Since these first exploratory works, microscopic studies based on different RPA models  have been used to study neutron-rich nuclei along the nuclear chart.  All neutron-rich nuclei seem to display a visible structure in the response to external fields at low energies which has been attributed to the pygmy resonance. Relativistic mean field models have also identified pygmy resonances. To cite a few of these works, we list Refs. \cite{Cat97,Vret01,Sar04,Pik06,Lian07,Paa07,Lit07,TL08,Bar08,Paa09,INY09,Lan09,Car10,Roc12,Vre12}. It is therefore clear that pygmy resonances constitute a new phenomenon in nuclear physics which appeared in the context of neutron-rich nuclei and the effect is also enhanced in nuclei close to the drip-line with small binding energies.

Recently a new and powerful method has been developed to study not only pygmy resonances but also nuclear large amplitude collective motion \cite{BR08,Bul11,Ste11,Bul13,Ste15,Bul16,Bulg19}. The framework relies on the time dependent superfluid local density approximation (TDSLDA).  This is an extension of the Density Functional Theory (DFT) to superfluid nuclei and can handle the response to external time-dependent fields. All degrees of freedom are taken into account  without any restrictions with all symmetries implemented such as translation, rotation, parity, local Galilean covariance, local gauge symmetry, isospin symmetry and minimal gauge coupling to electromagnetic fields \cite{Bulg19}. 

As in the time-dependent Hartree-Fock-Bogoliubov theory, the time evolution of the nucleus  is governed by the time-dependent mean field
\begin{eqnarray}
\lefteqn{i\hbar\frac{\partial}{\partial t} 
\left  ( \begin{array} {c}
  U({\bf r},t)\\  
  V({\bf r},t)\\ 
\end{array} \right ) \nonumber }\\
&& =
\left ( \begin{array}{cc}
h({\bf r},t)&\Delta({\bf r},t)\\
\Delta^*({\bf r},t)&-h^* ({\bf r},t)
\end{array} \right )  
\left  ( \begin{array} {c}
  U({\bf r},t)\\
  V({\bf r},t)\\ 
\end{array} \right ),
\end{eqnarray}
where $h(\mathbf{r},t)$ is the single-particle Hamiltonian and $\Delta({\bf r},t)$ is the pairing field. Both are obtained self-consistently from an
energy functional, i.e., a Skyrme interaction. The time-dependent external electromagnetic field $\bf A$  enters the hamiltonian by means of the minimal gauge coupling $\mbox{{\boldmath{$\nabla$}}}_{A} =\mbox{{\boldmath{$\nabla$}}} - i{\bf A}/{\hbar c} $.  The energy spectrum is obtained by a Fourier transformation of the time evolution of the nuclear density.

In Ref. \cite{Ste15} the first TDSLDA calculations were reported for relativistic Coulomb excitation in a collision of $^{238}$U + $^{238}$U.  The results show that a considerable amount of electromagnetic strength occurs at low energies, around $E_x \sim 7$ MeV. This additional structure was  attributed to  the excitation of the pygmy dipole resonance (PDR). 

\subsubsection{Electron Scattering}

For very forward electron scattering and small energy transfers, Siegert theorem \cite{Sie37,Sa51} allows one to show that the Coulomb and electric form factors appearing in electron scattering differential cross sections are proportional to each other and one obtains for electric multipole excitations \cite{Ber07}
\begin{equation}
\frac{d\sigma}{d\Omega dE_{\gamma}}=\sum_{L}\frac{dN_{e}^{(EL)}\left(
E_{e},E_{\gamma},\theta\right)  }{d\Omega dE_{\gamma}}\ \sigma_{\gamma}%
^{(EL)}\left(  E_{\gamma}\right)  , \label{EPA}%
\end{equation}
valid in the long-wavelength approximation, i.e.,  for excitation energies $ E_\gamma \ll \hbar c/R$, with $E_{e}$ being the electron energy, and the equivalent photon number given by \cite{Ber07}
\begin{eqnarray}
&&\frac{dN_{e}^{(EL)}\left(  E_{e},E_{\gamma},\theta\right)  }{d\Omega dE_{\gamma}}
  =\frac{4L}{L+1}\frac{\alpha}{E_{e}}\left[  \frac{2E_{e}}{E_{\gamma}}\sin\left(
\frac{\theta}{2}\right)  \right]  ^{2L-1}\nonumber\\
&  \times&\frac{\cos^{2}\left(  \theta/2\right)  \sin^{-3}\left(
\theta/2\right)  }{1+\left(  2E_{e}/M_{A}c^{2}\right)  \sin^{2}\left(
\theta/2\right)  }\nonumber \\
&\times& \left[  \frac{1}{2}+\left(  \frac{2E_{e}}{E_{\gamma}}\right)
^{2}\frac{L}{L+1}\sin^{2}\left(  \frac{\theta}{2}\right)  +\tan^{2}\left(
\frac{\theta}{2}\right)  \right]  . \label{EPAE}%
\end{eqnarray}

As with the case of Coulomb excitation, the cross section for electron scattering in this limit, and for large electron energies, are proportional to the cross sections for the electric multipolarity $EL$ induced by real photons. 

In both cases, this is a very useful relationship, as it is nearly impossible to study the interactions of real photons with radioactive nuclei in the laboratory. The conditions above are also hard to establish, except for electron-ion colliders, which have been promised in radioactive beam facilities, but not yet fullly realized \cite{Ant11,Sud12}.  

The excitation of the electric pygmy dipole resonance at large angles by inelastic electron scattering with existing facilities has been reported in this special issue by Pomomarev et al. \cite{Pon19}. They demonstrate that the excitation of pygmy resonance states in (e,e') reactions is predominantly of transversal character for large scattering angles. They were also able to extract the fine structure of the pygmy states at low excitation energies.

Electron scattering on halo nuclei was explored theoretically in Ref. \cite{Ber05} and a comparison with fixed-target experiments was  presented. For a given electron energy $E_{e}$, the total cross section for the dissociation of halo nuclei was shown to be 
\begin{equation}
\sigma_{e}(E_{e}) =64\sqrt{2} \pi  {e_{eff}^{2} \over \mu c^{2}S} \ln \left({E_{e} \over S}\right),
\end{equation}
where $\mu$ is the reduced mass of the halo nucleus, treated as a two-body cluster-like nucleus, and $e_{eff}$ is the effective charge involved in the transition.

For normal nuclei, with $S \simeq$ few MeV, the halo nucleus electron-disintegration cross section is negligible. The formula above predicts a dependence on the inverse of the separation energy. In an hypothetical situation, with  $S =100$ keV, $E_{e}=10$ MeV,  $e_{eff}= e$, and $\mu  = m_{N}\sim 10^3$ MeV, this equation yields a non-negligible 25 mb for the dissociation cross section. The  electro-disintegration cross section increases very slowly with the electron energy posing a challenge for future experiments. In contrast, it was shown in Ref. \cite{Ber05}, that at low electron energies the cross sections increase much faster with $E_{e}$. These conclusions are of relevance for the designing of experiments in future electron-ion colliders and to resolve the energy spectrum around the pygmy resonance.

\subsubsection{Pygmies Are Real}

Based on all evidences discussed above, we conclude that the existence of a PDR is verified in all theoretical methods involving two-body, three-body and many-body models dealing with the electromagnetic response of neutron-rich nuclei. It has also been abundantly observed in experiments. Next we discuss their relation to astrophysics.

\section{Equation of State of Neutron Stars} 

\subsection{Symmetry Energy}

The structure of neutron stars is dependent on the equation of state (EOS) of infinite neutron matter with  a small proton component \cite{LP04}.  Since most of the experimental data related to EOS is obtained with nuclei with $N\sim Z$, it is important to have an accurate description of the part of the EOS depending on the fraction $\delta = (N-Z)/A$, called by  {\it symmetry energy}.  

The energy per nucleon in nuclear matter can be expanded in a Taylor series around $N=Z$ yielding
 \begin{equation}
\epsilon(\rho,\delta) = \epsilon (\rho, 0) + S(\rho) \delta^{2}+\cdots,\label{epsilon}
\end{equation}
where here  $S$ denotes the symmetry energy. For the infinite matter in neutron stars one uses, $\delta = (\rho_{n}-\rho_{p})/\rho$, where $\rho=\rho_{p} + \rho_{n}$ is the total nuclear density. The incompressibility of nuclear matter is defined as 
  \begin{equation}
K_{0}= 9 \rho_{0}^{2}\left. {\partial^{2}\epsilon \over \partial \rho^{2}}\right|_{\rho_{0}},
\end{equation} 
and is perhaps the easiest quantity to relate infinite nuclear matter to experimentally observed properties such as monopole excitations in nuclei \cite{AB13}.

The function $S(\rho)$ can also be expanded around the saturation density of nuclear matter, $\rho_0\simeq 0.16$ fm$^{-3}$,
\begin{equation}
S(\rho) =J+{\L\over 3} {{\rho-\rho_{0}}\over \rho_{0}}+\cdots, \label{symme}
\end{equation}
where  the  bulk symmetry energy is given by $J=S(\rho_{0})$ and the slope parameter is given by 
\begin{equation}
L=3\rho_{0}\left.{dS(\rho)\over d\rho}\right|_{\rho_{0}}. 
\end{equation}
The binding energy per nucleon at saturation is $\epsilon(\rho_0,0)\simeq -16$ MeV. Fits to nuclear properties such as masses and excitation energies with microscopic models have indicated that $J\approx 30$ MeV \cite{Li14}. Because the theoretical models have been fitted to reproduce normal nuclei, it is not easy to find out which value of $L$ from these models is the proper one to use in the EOS of  neutron stars.

The EOS of homogeneous nuclear matter is a relation of the pressure and density, given by 
\begin{equation}
p(\rho,\delta)= \rho^{2}{d\epsilon(\rho,\delta) \over d\rho}. 
\end{equation}
It is strongly dependent on  $S$, because for  $\delta =1$, i.e., neutron matter, and $\rho \sim \rho_0$ one obtains $p = L\rho_0/3$. Therefore, the slope parameter $L$ is crucial to describe the EOS of neutron matter. 
Experimental and theoretical analyses of various kinds of nuclear structure and nuclear reactions show that $L$ is not well known, varying with the range $0$ and 150 MeV \cite{Roc11,Roc15}. The  explosion mechanism of core-collapse supernovae is also dependent on the EOS of nuclear matter and its characteristics out of neutron/proton symmetry \cite{LP04,Gle97,Web05,Hae07,LP12,BP12,Heb13}.

\subsection{TOV Equation}

The role of the pressure in obtaining the structure of neutron stars is determined by the solutions of the equations of hydrostatic equilibrium. They are supposed to be governed by  the Tolman-Oppenheimer-Volkoff (TOV) equations, an extension of Newton's laws including special relativity and general relativity corrections. The TOV equations are a set of coupled  differential equations of the form (here we use $c=1$)
\begin{eqnarray}
{dp \over dr} &=& -G {\rho (r) m(r) \over r^{2}} \left[ 1 + {p(r) \over \rho (r)}\right] \left [1 + {4\pi r^{3}p(r) \over m(r)}\right] \nonumber \\
&\times& \left[ 1-{2Gm(r)\over r}\right]^{-1}
\end{eqnarray}
\begin{eqnarray}
{dm(r) \over dr}=4\pi r^{2}\rho(r),
\end{eqnarray}
where $m(r)$ is the enclosed mass profile of the star up to radius $r$ and $G$ is the gravitational constant. The solutions of the TOV equations for a given EOS are able to predict the mass and radii of neutron stars. As reported in Ref. \cite{LP05}, there is a large variation among observational data and theories for masses of neutron stars. 

In order to solve the TOV equations one needs to know the relationship of pressure and energy density (EOS), $p(\rho)$ and as we discussed above, this relates very closely to the slope parameter $L$. And how much do we know about it? This is only part of the puzzle, but the one which might need most improvements.

\section{Slope Parameter and Pygmy Resonances}

\subsection{Skyrme Forces}

There exist various theoretical models to describe the bulk properties of nuclei, such as their masses and nucleon density distributions. A widely used model relies on  the  Hartree-Fock-Bogoliubov (HFB) theory using Skyrme interactions \cite{RS04}. The Skyrme forces  are contact interactions embodying  coordinate, spin and isospin dependence.  The energy density functional $\epsilon[\rho]$ is a straightforward byproduct of the model. As mentioned above, this density dependence is exactly what one needs to deduce the mass and radii of neutron stars using the TOV equations \cite{Gle97}.

There are hundreds of  Skyrme interactions that have been devised and that are able to describe successfully a limited number of nuclear properties. Some of them have recently been fitted by means of a computational description of global  masses and other nuclear properties \cite{Bert09}. This new era of intense developments in computational power has allowed for a better constraint of the interactions appropriate to fit global properties of nuclei. However, when extrapolated to densities below and above the saturation density, the equations of state $\epsilon(\rho)$ tend to diverge \cite{Brow00}. Therefore, there is a strong interest in the literature to pinpoint those Skyrme interactions that better describe neutron matter properties. A glimpse of this difficulty is shown in Table \ref{NMp} where we show some of the predictions of the Skyrme models for the input needed to neuron star properties. It is evident that, whereas $K_{0}$ and $J$ tend to agree among many of the predictions, the slope parameter $L$ is the least constrained. 

\begin{table}[ht]
\begin{center}
\caption{Properties of nuclear matter at the saturation density as predicted by some Skyrme models. All numbers are in units of MeV. 
The parameters for the Skyrme forces were taken from \cite{Bei75,Dob84,Rei95,Bro98,Cha98,Rei99,Gor05,Dut12}. For more details see Ref. \cite{BV19}
\label{NMp}}
\begin{tabular}{|c|c|c|c||c|c|c|c|c|}
\hline\hline
 & $K_{0}$ & $J$   &  $L$ &  & $K_{0}$ & $J$   &  $L$   \\ 
\hline 
SIII & 355. & 28.2 & 9.91& SLY5 & 230. & 32.0 & 48.2 \\
\hline 
SKP & 201. & 30.0 & 19.7& SKXS20 & 202. & 35.5 & 67.1\\
\hline
SKX & 271. & 31.1 & 33.2 & SKO & 223. & 31.9 & 79.1 \\
\hline
HFB9 & 231. &  30.0 & 39.9   & SKI5 & 255. & 36.6 & 129.      \\
\hline
\hline
\end{tabular}
\end{center}
\end{table}

\subsection{Dipole Polarizability}

Microscopic theories based on energy density functionals  with Skyrme forces or  relativistic models suggest that the nuclear dipole polarizability $\alpha_{D}$ defined as 
\begin{equation}
\alpha_{D} = {\hbar\over 2\pi^{2} e^{2}}\int {\sigma_{\gamma}^{E1} (E) \over E^{2}} dE = {8 \pi \over 9} \int {dB_{E1}(E)\over dE} {dE \over E}  
\end{equation}
is an additional quantity able to constrain the symmetry energy \cite{RN10,Tam11}.  These are easily extracted from Coulomb excitation experiments. The reason is simple: the virtual photon numbers entering Eq. \ref{dsde} have a $n_{E1}\sim \ln(1/E)$ dependence with energy, which together with the $1/E$ term favors the low energy part of the spectrum where pygmy resonances are located. Hence, a measurement of Coulomb dissociation (or electron scattering) is nearly proportional to the dipole polarizability. 

Experiments exploring  Coulomb excitation, and polarized proton scattering off neutron-rich nuclei to extract the dipole polarizability and its relation to the slope parameter have been reported in a several publications in Refs. \cite{Tam11,Adr05,Wie09,Sch08,Iwa12,Cha80,Rye02,End00,Pol12,Her97,Zil02,Vol06,Sav06,Sav08,End09,End10,Ton10,Hag16,Bir17,Ros13}, just to cite a few of them. To date, there is a large variation in the measurements of $\alpha_{D}$ in the order of 20-30\%. This is still too large to constrain most of the energy functionals stemming from Skyrme and relativistic models. Developments on nuclear reaction theory are also necessary to obtain the desired accuracy in the experimental analyses \cite{Bra16}.

A correlation between $\alpha_{D}$, the strength of pygmy resonances,  and the neutron skin in nuclei, 
\begin{equation}
\Delta r_{np} = \left<r_{n}^{2}\right>^{1/2}- \left<r_{p}^{2}\right>^{1/2},
\end{equation}
was also shown to exist in Refs. \cite{Car10,Piek11}. Several correlations have been found among neutron skin, dipole polarizability, and the slope parameter \cite{Brow00,Piek11,HP01,TB01,Fur02}. The neutron excess in a nucleus builds up a neutron pressure that is larger than the pressure due to the protons. The neutron pressure also contributes to the energy per nucleon which is a function of the nuclear density and its nucleon asymmetry. Therefore, neutron skins in nuclei are expected to be naturally correlated to the symmetry energy. 

Experiments dedicated to the measurement of neutron skins have also been the subject of large experimental interest, ranging from electron scattering (with the parity violation part of the interaction), anti-proton annihilation, and heavy ion collisions \cite{Piek11,Jast04,Aum17,Tam11,Ros13}.  An electron-ion collider directly using the electromagnetic interaction as a probe would be an ideal tool, but are still far from being fully realized with neutron-rich nuclei \cite{Ant11,Sud12}. It is possible that  the neutron skin can be accessed in fragmentation reactions using inverse kinematics, as proposed in Ref.  \cite{Aum17}. These are rather easy experiments using present radioactive beam facilities. The proposal is based on the measurement of neutron-changing cross sections by the detection of all  fragments with at least one neutron removed \cite{Aum17}. Another proposal suggests that a subtle but well-known phenomenon, the Maris effect, could be used to access information on the neutron skin \cite{SBA18}. These could be achieved with polarized proton targets in quasi-free (p,2p) reactions.

\section{Nucleosynthesis}
The physics cases discussed in the previous sections both for Coulomb excitation by heavy ions and also by electron scattering are useful to assess information on  radiative capture reactions in stars, such as (n,$\gamma$) and (p,$\gamma$) \cite{Gor98,SK02,Ang08,TL15,Tso15,TL19}.
So far, the pygmy resonances have been mainly probed using electromagnetic excitation (see, e.g., Eq. \ref{dsde}) from which one can extract the photodissociation cross section \ref{photo}. Using the detailed balance theorem, it was proposed in Ref. \cite{BBR86} that radiative capture reactions of relevance for astrophysics could be obtained in electromagnetic dissociation experiments. This was proved to be a very useful tool in numerous experiments, e.g., Refs. \cite{Iwa99,Mot02,Fle05,Hor06,Adel11,Alt14} and became a state of the art tool in nuclear accelerator facilities.

The impact of the pygmy resonances in nuclear astrophysics is also imprinted in the energy balance of the reactions in the stellar medium. It has been shown, and up to now little explored, that  the energy balance in the nuclear reaction networks  may change the nuclear abundances appreciably if one includes pygmy resonances \cite{Gor98}.

\section{Beyond Pygmy Resonances and Neutron Skins}

Experiments with nuclei on earth are very limited  by beam intensities, detection efficiencies, etc.  Guidance by theory is crucial for experimental success, but this symbiosis also has limitations. Besides, nuclei are not prototypes of neutron stars. Perhaps the closest examples of neutron stars on earth are neutron rich nuclei. But neutron stars are bound by gravity and not by the strong interaction. 

Theorists are limited by computational power and by the lack of knowledge of crucial parts of the physics ingredients necessary for their goals such as an accurate description of the interface of quarks, gluons and nucleon degrees of freedom. Based on the information collected in relativistic nuclear collisions, there is increasing evidence that a phrase transition exists between the quark-gluon phase and the nucleons in nuclei even at low temperatures. 

If there is a phase transition in the core of neutron stars due to a large density, then the symmetry energy obtained with nucleonic degrees of freedom is not enough to model the structure of neutron stars. There is a long way for experimentalists to improve their devices and theorists to refine their models of nuclear matter. Accumulated experience obtained in the last decades, tells us that every so often an idea such as the relation of spectra of pygmy resonances and of neutron skins to the symmetry energy help us to pave the way.

\section{Acknowledgements}
I am honored to have met Pier Francesco. I wish to thank him for his dedication to nuclear physics and the education of a new generation of nuclear scientists. Maybe I do this personally in another universe.

This work was supported in part by the U.S. DOE grant DE- FG02-08ER41533 and the U.S. National Science Foundation Grant No. 1415656.


\begin{thebibliography}{}
%
%
\bibitem{BK47}
G. C. Baldwin and G. S. Klaiber, Phys. Rev. 73 (1947) 1156; Phys. Rev. 71 (1948) 3.
\bibitem{BG37}
W. Bothe and W. Gentner, Z. Phys. 106 (1937) 236.
\bibitem{Mig44}
A. B. Migdal, J. Phys. USSR 8 (1944) 331.
\bibitem{BBB98}
P. F. Bortignon, A. Bracco and R. A. Broglia, {\it Giant Resonances, Contemporary Concepts in Physics} (Rutledge, 1998). 
\bibitem{BF75}
B. L. Berman and S. C. Fultz, Rev. Mod. Phys. 47 (1975) 713.
\bibitem{Ber77}
R. Bergere, {\it Lecture Notes in Physics} 61 (1977) 1; 108 (1979) 138.
\bibitem{Aum05}
T. Aumann, Eur. Phys. J. A 26 (2005) 441.
\bibitem{ABS95}
T. Aumann, C.A. Bertulani and K. Suemmerer, Phys. Rev. C 51 (1995) 416.
\bibitem{BV99} 
C.A. Bertulani and V. Ponomarev, Phys. Reports 321 (1999) 139.
\bibitem{Bar61}
G.A. Bartholomew, Annu. Rev. Nucl. Sci. 11 (1961) 259.
\bibitem{Brz69}
J. S. Brzosko,  et al., Can. J. Phys. 47 (1969) 2849.
\bibitem{MDB71}
R. Mohan, M. Danos, and L. C. Biedenharn, Phys. Rev. C 3 (1971) 1740.
\bibitem{MDB71}
Radhe Mohan, M. Danos and L. C. Biedenharn, Phys. Rev. C 3 (1971) 1740.
\bibitem{NK87}
 T. Nomura and S. Kubono. ``Soft giant resonance - Experimental proposal to the Japanese Hadron project (now J-PARC)'' (1987) June.
 \bibitem{Tan85}
I. Tanihata et al., Phys. Lett. B 160 (1985) 380; Phys. Rev. Lett. 55(1985) 2676; Phys. Lett. B 206 (1988) 592.
\bibitem{Kob89}
T. Kobayashi, et al., Phys. Lett. B 232 (1989) 51.
\bibitem{HJ87}
P.G. Hansen and B. Jonson, Europhys. Lett. 4  (1987) 409.
\bibitem{BB86}
C.A. Bertulani and G. Baur, Nucl. Phys. A 480 (1988) 615.
\bibitem{Ser47}
R. Serber, Phys. Rev. 72 (1947) 1008.
\bibitem{Orr92}
N.A. Orr, et al., Phys. Rev. Lett. 69 (1992) 2050.
\bibitem{Aum00}
T. Aumann et al., Phys. Rev. Lett. 84 (2000) 35.
\bibitem{BH04}
C.A. Bertulani and P.G. Hansen, Phys. Rev. C70 (2004) 034609. 
\bibitem{BM93}
C.A. Bertulani and K.W. McVoy, Phys. Rev. C 48 (1993) 2534. 
\bibitem{HBE96}
K. Hencken, G. Bertsch, H. Esbensen, Phys. Rev. C 54 (1996) 3043.
\bibitem{BH00}
F. Barranco and  P.G. Hansen, Eur. Phys. J. A 7 (2000) 479.
\bibitem{Mar02}
Jerome Margueron, Angela Bonaccorso and David M. Brink, Nucl. Phys. A 703 (2002) 105.
\bibitem{BCH93}
C.A. Bertulani, L.F. Canto and M.S. Hussein, Phys. Reports 226 (1993) 281.
\bibitem{Iek13}
K. Ieki, et al., Phys. Rev. Lett. 70 (1993) 730.
\bibitem{Sac13}
D. Sackett, et al., Phys. Rev. C 48 (1993) 118. 
\bibitem{BB88}
C.A. Bertulani and G. Baur, Phys. Reports 163 (1988) 299.
\bibitem{BBH91}
C. A. Bertulani, G. Baur and M. S. Hussein, Nucl. Phys. 526 (1991) 751.
\bibitem{BS92}
C.A. Bertulani and A. Sustich, Phys. Rev. C46 (1992) 2340.
\bibitem{Nak94}
T. Nakamura et al.,  Phys. Lett. B331 (1994) 296.
\bibitem{Shi95}
S. Shimoura et al., Phys. Lett. B 349 (1995) 29.
\bibitem{Ter91}
N. Teruya, C.A. Bertulani, S. Krewald, H. Dias and M. S. Hussein, Phys. Rev. C 43 (1991) 2049.
\bibitem{Ots94}
T. Otsuka, M. Ishihara, N. Fukunishi, T. Nakamura, M. Yokoyama, Phys. Rev. C 49 (1994) R2289.
\bibitem{Barr01}
F. Barranco, P.F. Bortignon, R.A. Broglia, G. Col\`o, and E. Vigezzi, Eur. Phys. J. A 11 (2001) 385.
\bibitem{Pot10}
G. Potel, F. Barranco, E. Vigezzi and R.A. Broglia,  Phys. Rev. Lett. 105 (2010) 172502.
\bibitem{Barr17}
F. Barranco, G. Potel,  R.A. Broglia, and E. Vigezzi,  Phys. Rev. Lett. 119 (2017) 082501.
\bibitem{Man19}
Manju, Jagjit Singh, Shubhchintak, and R. Chatterjee, Eur. Phys. J. A (2019) 55.
\bibitem{Brog19}
R.A. Broglia, F. Barranco, A. Idini, G. Potel and  E. Vigezzi, arXiv:1806.09409 (2019).
\bibitem{Barr19}
F. Barranco, G. Potel,  E. Vigezzi and R.A. Broglia,  arXiv:1812.01761 (2018).
\bibitem{Pon19}
V. Ponomarev, D. H. Jakubassa-Amundsen, A. Richter and J. Wambach, arXiv:1904.08772 (2019).
\bibitem{BC92}
C.A. Bertulani and L.F. Canto, Nucl. Phys. A 539 (1992) 163.
\bibitem{BB93}
G.F. Bertsch and C.A. Bertulani, Nucl. Phys. A 556 (1993) 136.
\bibitem{BS93}
P. Banerjee, R. Shyam, Nucl. Phys. A 561 (1993) 112.
\bibitem{Kid94}
T. Kido,  K. Yabana,  Y. Suzuki, Phys. Rev. C 50 (1994) R1276.
\bibitem{BS96}
P. Banerjee, R. Shyam, J. Phys. G, Nucl. Part. Phys. 22 (1996) L79.
\bibitem{Kid96}
T Kido, K Yabana, Y Suzuki, Phys. Rev. C 53 (1996) 2296.
\bibitem{Ban02}
P. Banerjee, G. Baur, K. Hencken, R. Shyam, D. Trautmann, Phys. Rev. C 65 (2002) 064602.
\bibitem{Mar03}
Jerome Margueron, Angela Bonaccorso and David M.Brink, Nuclear Physics A 720 (2003) 337.
\bibitem{IB05}
Awad A.Ibraheem and A. Bonaccorso, Nucl. Phys. A 748 (2005) 3.
\bibitem{Bert05}
C.A.Bertulani, Phys. Rev. Lett. 94 (2005) 072701.
\bibitem{TB05}
S. Typel, G. Baur, Nucl. Phys. A 759 (2005) 247.
\bibitem{Mor05}
A. M. Moro, F. Perez-Bernal, J. M. Arias, and J. Gomez-Camacho, Phys. Rev. C 73 (2006) 044612.
\bibitem{OB09}
K. Ogata  and C.A. Bertulani, Prog. Theor. Phys. 121 (2009) 1399.
\bibitem{Oga09}
K. Ogata, T. Matsumoto, Y. Iseri, and M. Yahiro, J. Phys. Soc. Jpn. 78 (2009) 084201.
\bibitem{OB10}
K. Ogata and C.A. Bertulani, Prog. Theor. Phys. 123 (2010) 701.
\bibitem{KB12}
Ravinder Kumar and Angela Bonaccorso, Phys. Rev. C 86 (2012) 061601.
\bibitem{HTS13}
K. Hagino, I.  Tanihata, H. Sagawa, {\it Exotic Nuclei Far from the Stability Line. 100 Years of Subatomic Physics} (2013) 231.
\bibitem{Shub15}
Shubhchintak, Neelam, R. Chatterjee, R. Shyam, K. Tsushima, Nucl. Phys. A 939 (2015) 101.
\bibitem{Ala17}
N. Alamanos, C.A. Bertulani, A.  Bonaccorso, A. Bracco, D. M. Brink, G. Casini, {\it Rewriting Nuclear Physics textbooks: 30 years with radioactive ion beam physics}, Eur. Phys. J. Plus 132 (2017) 37. 
\bibitem{BBFV16}
R. A. Broglia, P. F. Bortignon, F. Barranco, E. Vigezzi, A. Idini and G Potel, Phys. Scr. 91 (2016) 063012.
\bibitem{Cap18}
P. Capel, D. Phillips, H.-W. Hammer, Phys. Rrev. C 98 (2018) 034610.
\bibitem{PJZ96}
A. Pushkin, B. Jonson, and M. V. Zhukov, J. Phys. G 22 (1996) L95.
\bibitem{Zhu93}
M.V. Zhukov,  et al., Phys. Rep. 231  (1993) 151.
\bibitem{Dan98}
B.V. Danilin, J. Thompson, J.S. Vaagen, M.V. Zhukov, Nucl. Phys. A 632 (1998) 383.
\bibitem{Ber07}
C.A. Bertulani, Phys. Rev. C 75 (2007) 024606.
\bibitem{SIS90}
Y. Suzuki, K. Ikeda, and H. Sato, Prog. Theor. Phys. 83 (1990) 180.
\bibitem{VNW92}
 P. Van Isacker, M. A. Nagarajan, and D. D. Warner, Phys. Rev. C 45,  (1992) R13.
\bibitem{GT48}
M. Goldhaber and E. Teller, Phys. Rev. 74 (1948) 1046.
\bibitem{SJ50}
H. Steinwedel and J. H. D. Jensen, Z. Naturforschung 5A (1950) 413.
\bibitem{Mye77}
W. D. Myers, W. G. Swiatecki, T. Kodama, L. J. El-Jaick, and E. R. Hilf, Phys. Rev. C 15 (1977) 2032.
\bibitem{SG75}
S. Shlomo and G.F. Bertsch, Nucl. Phys. A243 (1975) 507.
\bibitem{RS04} 
P. Ring and P. Schuck, {\it The nuclear many-body problem}, Springer (2004).
\bibitem{Col13}
G. Colo, L. Cao, N.V. Giai, and L. Capelli, Comp. Phys. Comm. 184 (2013) 142.
\bibitem{FB90}
G. Bertsch and J. Foxwell, Phys. Rev. C 41 (1990) 1300.
\bibitem{Cat97}
F. Catara, E.G. Lanza, M.A. Nagarajan and A. Vitturi, Nucl. Phys. A 614  (1997) 86; Nucl. Phys. A 624 (1997)  449.
\bibitem{Vret01}
D. Vretenar, et al.,  Nucl.Phys. A 692 (2001) 496.
\bibitem{Sar04}
D. Sarchi, Phys.Lett. B601 (2004) 27.
\bibitem{Pik06}
J. Piekarewicz, Phys.Rev. C73 (2006) 044325.
\bibitem{Lian07}
Jun Liang, et al., Phys.Rev. C75 (2007) 054320.
\bibitem{Paa07}
N. Paar, D. Vretenar, E. Khan and G. Colo, Rep. Prog. Phys. 70 (2007) 691.
\bibitem{Lit07}
E. Litvinova, P. Ring and D. Vretenar, Phys. Lett. B 647 (2007) 111.
\bibitem{TL08}
N. Tsoneva and H. Lenske, Phys. Rev. C 77 (2008) 024321.
\bibitem{Bar08}
C. Barbieri, E. Caurier, K. Langanke, and G. Martínez-Pinedo, Phys. Rev. C 77 (2008) 024304.
\bibitem{Paa09}
N. Paar, Y. F. Niu, D. Vretenar, and J. Meng, Phys. Rev. Lett. 103 (2009) 032502.
\bibitem{INY09}
T. Inakura, T. Nakatsukasa, and K. Yabana, Phys. Rev. C 80 (2009) 044301.
\bibitem{Lan09}
E. G. Lanza, F. Catara, D. Gambacurta, M.V. Andres  and Ph. Chomaz, Phys. Rev. C (2009) 79 054615.
\bibitem{Car10}
A. Carbone, et al., Phys. Rev. C 81 (2010) 041301(R).
\bibitem{Roc12}
X. Roca-Maza, G. Pozzi, M. Brenna, K. Mizuyama, and G. Colo, Phys. Rev. C 85 (2012) 024601.
\bibitem{Vre12}
D. Vretenar, Y. F. Niu, N. Paar, and J. Meng, Phys. Rev. C 85 (2012) 044317.
\bibitem{BR08}
A. Bulgac and K. J. Roche, J. Phys. Conf. Ser. 125 (2008) 012064.
\bibitem{Bul11}
A. Bulgac, Y.-L. Luo, P. Magierski, K. J. Roche, and Y. Yu, Science 332 (2011) 1288.
\bibitem{Ste11}
I. Stetcu, A. Bulgac, P. Magierski, and K. J. Roche, Phys. Rev. C 84, (2011) 051309(R).
\bibitem{Bul13}
A. Bulgac, Annu. Rev. Nucl. Part. Sci. 63 (2013) 97.
\bibitem{Ste15}
I. Stetcu, C. Bertulani, A. Bulgac, P. Magierski,  and K.J. Roche, Phys. Rev. Lett. 114 (2015) 012701.
\bibitem{Bul16}
A. Bulgac, P. Magierski, K.J. Roche, and I. Stetcu,   Phys. Rev. Lett. 116 (2016) 122504.
\bibitem{Bulg19}
A. Bulgac, Physica Status Solidi B (2019) 1800592.
\bibitem {Sie37}
A.J.F. Siegert, Phys. Rev. 52 (1937) 787.
\bibitem {Sa51}
R.G. Sachs and N. Austern, Phys. Rev. 81 (1951) 705.
\bibitem{Ant11}
A.N. Antonov et al., Nuc. Inst. Meth. Phys. Res. A 637 (2011) 60. 
\bibitem{Sud12}
Toshimi Suda, et al., Prog. Theor. Exp. Phys. 1 (2012) 03C008.
\bibitem{Ber05}
C.A. Bertulani, Phys. Lett. B 585 (2004) 35.
\bibitem{LP04}
J. M. Lattimer and M. Prakash, Science 304 (2004) 536.
\bibitem{AB13}
P. Avogadro and C. A. Bertulani, Phys. Rev C 88 (2013) 044319. 
\bibitem{Li14}
{\it Topical Issue on Nuclear Symmetry Energy}, edited by Bao-An Li, A. Ramos, G. Verde, and I. Vida\~na, Eur. Phys. J. A 50 (2014) 2.
\bibitem{Roc11}  
X. Roca-Maza, M. Centelles, X. Vinas, and M. Warda, Phys. Rev. Lett. 106 (2011) 252501.
\bibitem{Roc15} 
X. Roca-Maza, X. Vinas, M. Centelles, B. K. Agrawal, G. Colo, N. Paar, J. Piekarewicz, and D. Vretenar, Phys. Rev. C 92 (2015) 064304.
\bibitem{Gle97} 
N.K. Glendenning,  {\it Compact stars: Nuclear physics, particle physics, and general relativity}, Springer, New York, (1997).
\bibitem{Web05} 
F. Weber, Prog. Part. Nucl. Phys. 54 (2005) 193.
\bibitem{Hae07} 
P. Haensel, A. Potekhin, and D. Yakovlev, {\it Neutron stars 1: Equation of state and structure}, Springer, New York (2007).
\bibitem{LP12} 
J.M. Lattimer, Annu. Rev. Nucl. Part. Sci. 62 (2012) 485.
\bibitem{BP12}  
{\it Neutron Star Crust}, editors C.A. Bertulani and J. Pieckarewicz, Nova Science Publishers,  New York (2012).
\bibitem{Heb13} 
K. Hebeler, J. M. Lattimer, C. J. Pethick, and A.~Schwenk, Astrophys. J. 773 (2013) 11.
\bibitem{LP05}
J.M. Lattimer and M. Prakash, Phys. Rev. Lett. 94 (2005) 111101.
\bibitem{Bert09}
G.F. Bertsch, C.A. Bertulani, W. Nazarewicz, N. Schunck and M.V. Stoitsov, Phys. Rev. C 79 (2009) 0343306.
\bibitem{Brow00}
B. Alex Brown, 85 (2000) 5296.
\bibitem{Bei75} 
M. Beiner et al., Nucl. Phys. A238 (1975) 29.
\bibitem{Dob84} 
J. Dobaczewski, et al., NPA 422 (1984) 103.
\bibitem{Rei95} 
P.-G. Reinhard, H. Flocard, Nucl. Phys. 584 (1995) 467.
\bibitem{Bro98} 
B. A. Brown, Phys. Rev. C58 (1998) 220.
\bibitem{Cha98} 
E. Chabanat, P. Bonche, P. Haensel, J. Meyer, and R. Schaeffer, Nucl. Phys. A635 (1998) 231.
\bibitem{Rei99} 
P.-G. Reinhard,et al., Phys. Rev. C 60 (1999) 14316.
\bibitem{Gor05} 
S. Goriely, et al., Nucl. Phys. A 750 (2005) 425.
\bibitem{Dut12} 
M. Dutra, O. Lourenco, J.S. Sa Martins, A. Delfino, J. R. Stone, and P. D. Stevenson, Phys. Rev. C 85 (2012) 035201.
\bibitem{BV19}
C.A. Bertulani and J. Valencia, arXiv:1904.01078 (2019).
\bibitem{RN10}
P.-G. Reinhard and W. Nazarewicz, Phys. Rev. C 81 (2010) 051303(R).
\bibitem{Tam11}
A. Tamii et al., Phys. Rev. Lett. 107 (2011) 062502.
\bibitem{Adr05}
P. Adrich et al., Phys. Rev. Lett. 95 (2005) 132501.
\bibitem{Wie09}
O. Wieland et al., Phys. Rev. Lett. 102 (2009) 092502.
\bibitem{Sch08}
R. Schwengner et al., Phys. Rev. C 78 (2008) 064314.
\bibitem{Iwa12}
C. Iwamoto et al., Phys. Rev. Lett. 108 (2012) 262501.
\bibitem{Cha80}
T. Chapuran, R. Vodhanel, and M. K. Brussel, Phys. Rev. C 22 (1980) 1420.
\bibitem{Rye02}
N. Ryezayeva et al., Phys. Rev. Lett. 89 (2002) 272502.
\bibitem{End00}
J. Enders et al., Phys. Lett. B 486, 279 (2000); Nucl. Phys. A 724 (2003) 243.
\bibitem{Pol12}
I. Poltoratska et al., Phys. Rev. C 85 (2012) 041304(R).
\bibitem{Her97}
R.-D. Herzberg et al., Phys. Lett. B 390, 49 (1997); Phys. Rev. C 60 (1999) 051307.
\bibitem{Zil02}
[A. Zilges et al., Phys. Lett. B 542 (2002) 43; A. Zilges, Nucl. Phys. A 731 (2004) 249 .
\bibitem{Vol06}
S. Volz et al., Nucl. Phys. A 779 (2006) 1.
\bibitem{Sav06}
D. Savran et al., Phys. Rev. Lett. 97 (2006) 172502.
\bibitem{Sav08}
D. Savran et al., Phys. Rev. Lett. 100 (2008) 232501; Phys. Rev. C 84 (2011) 024326.
\bibitem{End09}
J. Endres et al., Phys. Rev. C 80 (2009) 034302.
\bibitem{End10}
J. Endres et al., Phys. Rev. Lett. 105 (2010) 212503.
\bibitem{Ton10}
A. P. Tonchev et al., Phys. Rev. Lett. 104 (2010) 072501.
\bibitem{Hag16}
G. Hagen, et al., Nature Phys. 12 (2016) 186.
\bibitem{Bir17}
J. Birkhan et al., Phys. Rev. Lett 118 (2017) 252501.
\bibitem{Ros13}
D. Rossi et al. Phys. Rev. Lett. 111 (2013) 242503.
\bibitem{Bra16}
N. Brady, T. Aumann, C.A. Bertulani, J. Thomas, Phys. Lett. B 757 (2016) 553.
\bibitem{Piek11}
J. Piekarewicz, Phys. Rev. C 83 (2011) 034319.
\bibitem{HP01}
C. J. Horowitz and J. Piekarewicz, Phys. Rev. Lett. 86 (2011) 5647.
\bibitem{TB01} 
S. Typel and B. A Brown, Phys. Rev. C 64 (2011) 027302.
\bibitem{Fur02} 
R. J. Furnstahl, Nucl. Phys. A 706 (2002) 85.
\bibitem{Jast04} J. Jastrzebski, A. Trzcinska, P. Lubinski, B. Clos, F. J. Hartmann, T. von Egidy, and S. Wycech, Int. Journ. Mod. Phys. E 13, 343 (2004).
\bibitem{Aum17}
T. Aumann, C.A. Bertulani, F. Schindler, S. Typel, Phys. Rev. Lett. 119, (2017) 262501. 
\bibitem{SBA18}
Shubchintak, C.A. Bertulani, T. Aumann Phys. Lett. B 778 (2018) 30.
\bibitem{Gor98}
S. Goriely, S. Phys.Lett. B436 (1998) 10.
\bibitem{SK02}
S. Goriely and E. Khan,  Nucl. Phys. A706 (2002) 217.
\bibitem{Ang08} 
Christoper T. Angell, Ph.D. Thesis, submitted to the University of North Carolina (2008).
\bibitem{TL15}
Nadia Tsoneva, Horst Lenske, Nature Commun. 6 (2015) 7384; EPJ Web Conf. 93 (2015) 01021.
\bibitem{Tso15}
N. Tsoneva, S. Goriely, H. Lenske, and R. Schwengner, Phys. Rev. C 91 (2015) 044318.
\bibitem{TL19}
Nadia Tsoneva, Horst Lenske, arXiv:1410.2458 (2019).
\bibitem{BBR86}
G. Baur, C.A. Bertulani and H. Rebel, Nucl. Phys. A458 (1986) 188. 
\bibitem{Iwa99}
N. Iwasa, et al.,Phys. Rev. Lett. 83 (1999) 2910.
\bibitem{Mot02}
T. Motobayashi, Eur. Phys. J. A 13  (2002) 207.
\bibitem{Fle05}
F. Fleurot, et al., Phys. Lett. B 615 (2005) 167.
\bibitem{Hor06}
A. Horvath, et al., Eur. Phys. J. A 27, Supp. 1 (2006) 217.
\bibitem{Adel11}
E.G. Adelberger et al., Rev. Mod. Phys. 83 (2011) 195.
\bibitem{Alt14}
S.G. Altstadt, et al., Nuclear Data Sheets 120 (2014) 197.




\end{thebibliography}
\end{document}